\documentclass[twocolumn,showpacs,preprintnumbers,amsmath,amssymb,prb,aps]{revtex4-1}
\usepackage{graphicx}
\usepackage{dcolumn}
\usepackage{bm}
\bibpunct{}{}{,}{s}{}{}
\begin{document}

\preprint{14-July-11}

\title{Influence of the heterointerface sharpness on exciton recombination dynamics in an ensemble of (In,Al)As/AlAs quantum dots with indirect band-gap}

\author{T. S. Shamirzaev$^{1}$, J. Debus$^{2}$, D. S. Abramkin$^{1}$, D. Dunker$^2$, D.~R.~Yakovlev$^{2,3}$, D.~V.~Dmitriev$^{1}$, A.~K.~Gutakovskii$^{1}$, L. S. Braginsky$^{1}$, K. S. Zhuravlev$^1$, and M. Bayer$^2$}
\affiliation{$^1$A. V. Rzhanov Institute of Semiconductor Physics, Siberian branch of the Russian Academy of Sciences, 630090 Novosibirsk, Russia}
\affiliation{$^2$Experimentelle Physik 2, Technische Universit\"at Dortmund, 44227 Dortmund, Germany}
\affiliation{$^3$A. F. Ioffe Physical-Technical Institute, Russian Academy of Sciences, 194021 St. Petersburg, Russia}

\begin{abstract}
The dynamics of exciton recombination in an ensemble of indirect band-gap (In,Al)As/AlAs quantum dots with type-I band alignment is studied. The lifetime of confined excitons which are indirect in momentum-space is mainly influenced by the sharpness of the heterointerface between the (In,Al)As quantum dot and the AlAs barrier matrix. Time-resolved photoluminescence experiments and theoretical model calculations reveal a strong dependence of the exciton lifetime on the thickness of the interface diffusion layer. The lifetime of excitons with a particular optical transition energy varies because this energy is obtained for quantum dots differing in size, shape and composition. The different exciton lifetimes, which result in photoluminescence with non-exponential decay obeying a power-law function, can be described by a phenomenological distribution function $G(\tau)$, which allows one to explain the photoluminescence decay with one fitting parameter only.
\end{abstract}


\maketitle

\section{Introduction}

The kinetics of exciton recombination in semiconductor QDs is often analyzed in terms of an exponential decay with one characteristic recombination time.~\cite{Driell} However, the luminescence decay in QDs is typically nonexponential~\cite{Driell, Berstermann, Driel}, for which there are several reasons such as the contribution of dark excitons to the emission or the influence of Coulomb correlation effects. Nevertheless, for a single QD in the strong confinement regime the bright exciton PL is found to decay monoexponentially.~\cite{Zwiller,Berstermann} For an ensemble of such QDs, on the other hand, nonexponential decays are often found, and a statistical analysis of the time-resolved emission demonstrates that this behavior can be attributed to a dispersion of radiative and/or nonradiative lifetimes of QD confined excitons.~\cite{Driel,Lee}  This ensemble decay at a specific energy results from the superposition of monoexponential PL decays of excitons which are localized in QDs having different sizes, shapes, and compositions.~\cite{Bartel} In case of continuously distributed lifetimes $\tau$ of excitons, which are characterized by the same recombination energy, their PL decay can be described by a distribution function $G(\tau)$.

We demonstrated recently that a nonexponential long-time decay of the exciton PL is characteristic of indirect band-gap (In,Al)As/AlAs QDs with type-I band alignment.~\cite{Shamirzaeva,Shamirzaevb} In these structures the conduction band minimum is around the X-valley, while the valence band maximum is around the $\Gamma$-point. In QDs the momentum is no longer a good quantum number, but the wave function is distributed in momentum space over a range of ${\bf k}$-vectors that is inversely proportional to the quantum dot size. This extension is, however, still much smaller than the separation between the $\Gamma$- and the X-valley, so that the indirect character of the band gap is maintained. As a result, direct band-to-band transitions of electrons resulting in an emission of a photon are strongly suppressed. Instead, the radiative recombination requires the involvement of scattering by phonons or at the heterointerface, as has been demonstrated in indirect band-gap GaAs/AlAs and InAs/AlAs quantum wells.~\cite{Shamirzaevc, Braginsky} For (In,Al)As/AlAs QDs it has been ascertained that the radiative exciton recombination is mainly caused by the scattering at the heterointerface between the (In,Al)As QD and AlAs matrix.~\cite{Shamirzaevd,Shamirzaevg} Hence, the exciton recombination dynamics, namely the recombination time $\tau$ and the lifetime distribution $G(\tau)$ can yield valuable information on this interface.

In this paper the dynamics of the exciton recombination in ensembles of indirect band-gap (In,Al)As/AlAs QDs with varying sharpness of the QD/matrix interface is studied by time-resolved PL. We demonstrate that the radiative lifetime of the indirect in momentum-space exciton is strongly influenced by this sharpness. The decay can be well described by a power-law function $I(t) \sim (1/t)^{\alpha}$, which can be accounted by a phenomenological distribution function $G(\tau)$ based on a single fitting parameter.

\section{Samples and Experiment}
The studied self-assembled (In,Al)As QDs, embedded in an AlAs matrix, were grown by molecular-beam epitaxy (Riber-32P system) on semi-insulating (001)-oriented GaAs substrates. The structures have one QD sheet sandwiched between 50-nm-thick AlAs layers grown on top of a 200-nm-thick GaAs buffer layer. The nominal amount of deposited InAs was about 2.5 monolayers. A 20-nm-thick GaAs cap layer protects the AlAs layer against oxidation.

\begin{table*}
\caption{\label{table} Growth parameters and annealing temperatures for the studied (In,Al)As/AlAs QDs. The average diameter ($D_{\text{AV}}$), diameters corresponding to the smaller ($D_{\text{S}}$) and larger ($D_{\text{L}}$) half-width of the QD size distribution, the QD density, the size dispersion ($S_{\text{D}}$), and the composition of the (In,Al)As/AlAs QDs are given as well. Additionally, the exponent of the PL decay curve $\alpha$, and the parameter $\tau_0$ in the exciton lifetime distribution described by Eq.(\ref{eq3}) are listed. Note, the relation $\gamma=\alpha+1$ defines another parameter of the exciton lifetime distribution $\gamma$. For the structure S2 the diameter corresponding to the larger half-maximum in the QD size distribution is related to QDs with direct band gap. In order to have for structure S2 an additional ensemble of indirect band gap QDs with a characteristic diameter being different from $D_{\text{AV}}$ and $D_{\text{S}}$, we choose the diameter (marked in Fig.~\ref{fig1} by $D^{*}_{\text{L}}$ = 17~nm) that is related to indirect band gap QDs. The parameter values marked in the Table with an asterisk belong to the QD ensemble with the characteristic diameter $D^{*}_{\text{L}}$.}

\begin{tabular}{|c|c|c|c|c|c|c|c|c|c|c|c|c|c|c|} \hline
Structure & $T_{\text{g}}$/$t_{\text{GI}}$ & $T_{\text{an}}$ & $D_{\text{S}}$ &$D_{\text{AV}}$ & $D_{\text{L}}$ & QD & $S_{\text{D}}$ & Average & \multicolumn{3}{c|}{$\tau_{0}$} & \multicolumn{3}{c|}{$\alpha$}  \\
 & $^\circ$C/s & $^\circ$C & nm & nm & nm & density & \% & fraction & \multicolumn{3}{c|}{ns} & \multicolumn{3}{c|}{} \\
 & & & & & & $\times 10^{10}$ & & of InAs & \multicolumn{3}{c|}{} & \multicolumn{3}{c|}{}\\ \cline{10-15}
 & & & & & & cm$^{-2}$ & & in QDs & $I_{1/2}$ & $I_{\text{max}}$ & $I_{1/2}$ &  $I_{1/2}$ & $I_{\text{max}}$ &  $I_{1/2}$ \\
 & & & & & &  & & & $(D_{\text{S}})$ &($D_{\text{AV}})$ & ($D_{\text{L}})$ & ($D_{\text{S}}$) &($D_{\text{AV}}$) & ($D_{\text{L}}$) \\
 & & & & & &  & & &  & &  &  & &  \\ \hline
 S1 & 450/10 & - & 4.3 & 5.5 $\pm$ 0.21 & 7 & 10 & 40 & 0.99 & 130 & 100 & 70 & 1.75 & 1.75 & 1.30 \\
 S2 & 460/60 & - & 9 & 13.8$\pm$0.22 & 17* & 8.5 & 60 & 0.80 & 240 & 130 & 60* & 1.95 & 1.55 & 1.25* \\
 S3 & 510/60 & 700 & 15 & 18.3$\pm$0.15 & 22 & 4.2 & 52 & 0.47 & 2300 & 2100 & 700 & 1.75 & 1.50 & 1.35 \\
 S4 & 460/60 & 800 & 12 & 19.6$\pm$0.16 & 28 & 8.5 & 75 & 0.35 & 5400 & 5200 & 4000 & 2.45 & 2.40 & 2.08 \\ \hline
\end{tabular}

\end{table*}

Recently, we demonstrated that diameter, density, and composition of (In,Al)As/AlAs QDs are determined by the growth conditions such as substrate temperature $T_{\text{g}}$, and growth interruption time $t_{\text{GI}}$.~\cite{Shamirzaevd} Three structures S1, S2, and S3 were grown for this study using the conditions listed in Table~\ref{table}. According to these conditions the structures have different (In,Al)As QD alloy compositions.~\cite{Shamirzaevd} However, as demonstrated repeatedly, despite of the intermixing on the QD composition during the epitaxy, as-grown self-assembled QDs have a sharp QD/matrix interface.~\cite{Liu, Offermans} The reason for the sharp interface formation arises from the Stranski-Krastanov growth mode. The QD composition is determined by the intermixing during the dot formation due to mass transfer along the wetting layer.~\cite{Placidi,Panat,Heyn} The interfaces of QDs being independent on their composition are given by stable crystallographic planes, which ensure minimization of the interface energy. These planes provide interface stability against the intermixing with the matrix material during the overgrowth of the QDs. Nevertheless, the sharpness of the (In,Al)As/AlAs interfaces can be smeared out by means of high-temperature post-growth thermal annealing.~\cite{Shamirzaeve} Two of the studied structures, S2 and S3, were annealed for one minute at elevated temperatures $T_{\text{an}}$. Data on $T_{\text{an}}$ are listed in Table~\ref{table}, whereas technical details can be found in Ref.~\onlinecite{Shamirzaeve}.
The following, we will refer the annealed S2 structure as the S4 structure.
 
\begin{figure}[t]
\centering \includegraphics[width=7cm]{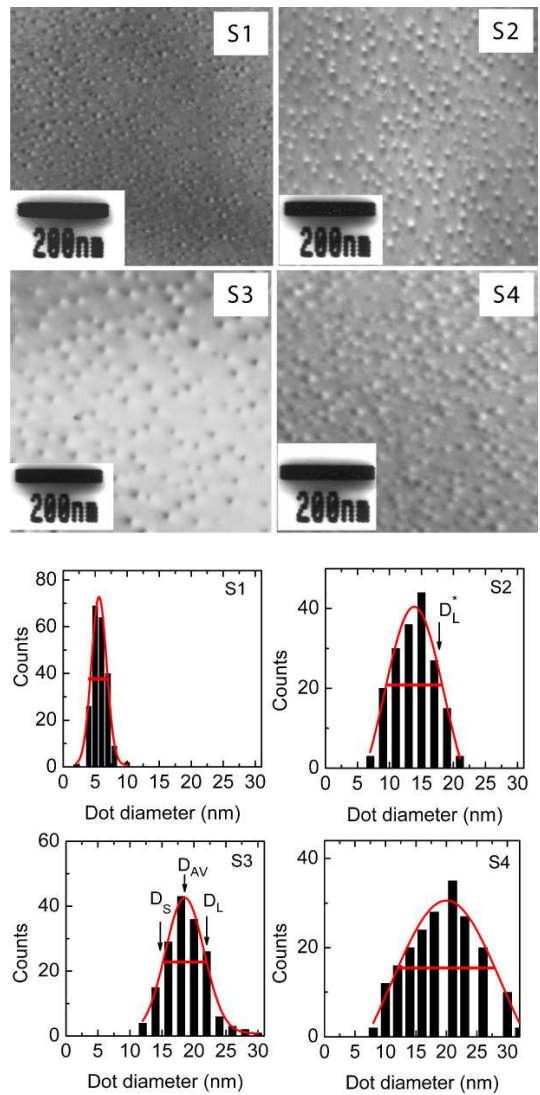}
\caption{\label{fig1} (Color online) TEM plane-view images (upper panels) and histograms of the QD-diameter distribution (lower panels), including the size dispersion fitted by Gaussian curves for the structures S1-S4. The histograms show the incidences of finding a QD with a specific diameter in a TEM image within an area ranging from 0.3 to 0.5~$\mu$m$^2$. The half-widths of the QD-size distribution are marked by the horizontal lines. The average diameter $D_{\text{AV}}$ and the large $D_{\text{L}}$ and small $D_{\text{S}}$ half-widths of the QD-size distribution are indicated for the structure S3. For the structure S2, $D^{*}_{\text{L}}$ is taken as the diameter corresponding to indirect band gap QDs, see caption to the Table~\ref{table}.}
\end{figure}

The QD size and density are studied by transmission electron microscopy (TEM) using a JEM-4000EX system operated at 250 keV acceleration voltage. From the TEM-images we find that the self-assembled (In,Al)As QDs are lens-shaped with a typical aspect ratio (height to diameter) of 1:4.~\cite{Shamirzaevb} TEM plane-view images and the respective histograms of the QD diameter distribution are shown in Fig.~\ref{fig1} for all structures. The average diameters $D_{\text{AV}}$ and the diameters corresponding to the larger $D_{\text{L}}$ and smaller $D_{\text{S}}$ half-widths of the QD size distribution obtained from the TEM images, are summarized in Table~\ref{table}. Additionally, the size dispersion $S_{\text{D}}$ is listed, which is defined by the ratio of the half-width of the Gaussian distribution of QD sizes to the average diameter  $S_{\text{D}} = 100 {\%} \times (D_{\text{L}} - D_{\text{S}})/ D_{\text{AV}}$.

The unannealed structure S1 has a relatively narrow distribution ($S_{\text{D}} =40{\%}$) around a small QD size of $D_{\text{AV}}=5.5$~nm. The structures S2, S3, and S4  contain QDs with larger diameters, which are also distributed over broader ranges. The largest lateral QD sizes are found for structure S4, which was annealed at the highest temperature of 800$^\circ$C. The comparison between the annealed structure S4 and the unannealed structure S2 indicates that post-growth annealing leads to an increase of the QD diameter with a broader distribution, in good agreement with previous results.~\cite{Shamirzaeve} The annealing results in an increase of the average QD diameter from 13.8 to 19.6~nm, and $S_{\text{D}}$ from 60{\%} to 75{\%}, respectively. The diameter increase by annealing is a result of InAs diffusion from the QD into the surrounding AlAs matrix. Therefore, the annealing results in the appearance of a diffused In$_{x}$Al$_{1-x}$As layer around the QD/matrix interface. It is obvious, that the thickness of this In$_{x}$Al$_{1-x}$As layer depends on the annealing temperature and duration. The sharpness of the interface is defined as the degree of the spatial separation between the different materials of (In,Al)As and AlAs.
Thus, the sharpness of the QD/matrix interface can be described in terms of the thickness of this diffused layer: a sharp (blurred) interface corresponds to a thin (thick) In$_{x}$Al$_{1-x}$As layer. Therefore, the structures S1-S4 provide us with a representative set of QD ensembles having different diameters and interface sharpness.

The steady-state and time-resolved PL measurements were performed at a temperature of $T=5$~K. For the excitation of the steady-state PL a He-Cd laser with a photon energy of 3.81~eV was used. The time-resolved PL experiments were established by the third harmonic of a Q-switched Nd:YVO$_4$ laser (3.49~eV) with a pulse duration of 5~ns. The pulse-repetition frequency was varied from 1 to 100~kHz and the pulse energy density was chosen between 40~nJ/cm$^2$ to 12~$\mu$J/cm$^2$. The emitted light was dispersed by a 0.5~m monochromator and detected by a GaAs photomultiplier operating in the time-correlated photon-counting mode. In order to monitor the PL decay in a wide temporal range up to 0.5~ms the time resolution of the detection system was varied between 1.6~ns and 200~ns.

\section{Experimental Results}

\subsection{Steady-state photoluminescence}
\begin{figure}[t]
\centering
\includegraphics[width=5cm]{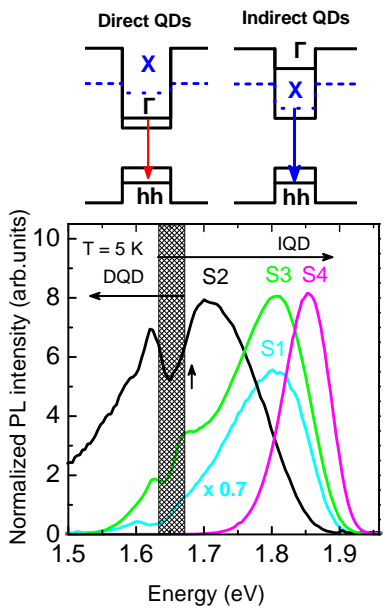}
\caption{\label{fig2} (Color online) Normalized PL spectra of the different (In,Al)As/AlAs QD structures S1 to S4, excited by a He-Cd laser with a power density of 10~W/cm$^2$. The PL intensity of the S1 structure is multiplied by a factor 0.7 for better visibility. The vertical dotted area marks the boundary between the emission from QD excitons either direct or indirect in momentum-space as calculated in Refs.~\onlinecite{Shamirzaevd},~\onlinecite{Shamirzaevb}. The vertical arrow indicates the energy of exciton recombination in QDs with typical diameter $D^{*}_{\text{L}}$ for the structure S2 (see caption of Table~\ref{table}). Schemes on top: Real space band structures of direct and indirect band gap QDs, including the energetically lowest $\Gamma$- and X-conduction band states as well as the heavy-hole (hh) states.}
\end{figure}

Normalized PL spectra of the studied structures are shown in Fig.~\ref{fig2}. Two PL bands, denoted by DQD and IQD, are observed in the spectra. Recently, we have shown that the PL kinetics of the DQD and IQD bands are considerably different. The DQD and IQD bands result from exciton recombination in QDs with direct and indirect band gap, respectively. Hereby, it has been evidenced by pulsed excitation measurements that the intensity of the DQD emission band drops quickly with a decay time shorter than 20~ns, while the IQD band decays nonexponentially during hundreds of $\mu$s.~\cite{Shamirzaevb} These two types of QDs coexist in these ensembles of (In,Al)As/AlAs QDs.~\cite{Shamirzaevb, Shamirzaevd} For the PL spectra of the S1, S2, and S3 structures the direct DQD band has been observed at energies below 1.65~eV. This energy well coincides with the observed and calculated boundary between the direct and indirect band-gap (In,Al)As/AlAs QDs.~\cite{Shamirzaevb, Shamirzaevd}

Since the shape of the PL emission reflects the distribution of QD sizes, we establish in the following the relation between the parameters characterizing the spectra and the geometrical quantities of the average diameter $D_{\text{AV}}$ and size dispersion $S_{\text{D}}$. This relation is given by the following features:
\begin{itemize}
\item The increase of $D_{\text{AV}}$ and $S_{\text{D}}$ for the as-grown structures S1 and S2 leads to a low-energy shift from 1.8 to 1.7~eV and an increase in full width at half maximum (FWHM) of the IQD band from 115 to 195~meV, respectively. Additionally, the intensity of the DQD band is increased by one order of magnitude in S2 compared to S1.
\item Increase in $D_{\text{AV}}$ and $S_{\text{D}}$, as a result of
high temperature annealing of the S2 structure and its
transformation to the S4 structure, leads for IQD band to high-energy
shift from 1.7 to 1.85~eV and decrease of FWHM from 195 to 75~meV.
The annealing results also in decrease of the DQD band intensity in
the  S3 structure and disappearance of such band in the  S4
structure.
\end{itemize}

In order to explain the obtained results we have to take into account that the energy of the optical transition is determined by two factors: (i) the quantum confinement energy which decreases with increasing QD size; and (ii) the band-gap energy of the (In,Al)As alloy in the QD which increases with decreasing InAs fraction. In lens-shaped QDs the confinement energy is mainly determined by the QD height. The average QD composition can be evaluated from comparison of the IQD band energy position with results from model calculations.~\cite{note} The determined QD compositions are collected in Table~\ref{table}.

The comparison of the observed optical transition energy with calculated ones shows that the low-energy shift and broadening of the IQD band, going from the S1 to the S2 structure, are caused by a decrease in quantum confinement and an increase in the size dispersion $S_{\text{D}}$, while the change in the QD composition is negligible. On the other hand, the high-energy shift of the IQD band for the S4 structure compared to S2 is due to an increase of the (In,Al)As alloy band-gap energy with decreasing InAs fraction in the QD alloy composition. This compensates the decrease in quantum confinement energy due to the annealing induced increase of the QDs height based on a fixed aspect ratio. The at first sight unusual reduction of the FWHM of the IQD band with increasing $S_{\text{D}}$ was explained in our previous study.~\cite{Shamirzaeve}

We calculated in Refs.~\onlinecite{Shamirzaevd},~\onlinecite{Shamirzaevb} the energy separation between the optical transitions of direct and indirect excitons and showed that it weakly depends on the QD size, shape, and composition. Therefore, the change in the relative intensity of the DQD band reflects the change in the relative layer density of the direct band gap QDs. This density is a function of the QD size and composition. The disappearance of the DQD band in the emission of the S4 structure results from the lower InAs fraction in the QD alloy composition. This, in turn, gives rise to a conversion of the band gap from a direct to an indirect one.

\subsection{Time-resolved photoluminescence}
For direct band gap (In,Ga)As QDs with typical exciton lifetimes of about 1~ns the condition that the QD exciton population does not exceed one exciton can be easily established. As an example, for (In,Ga)As/GaAs QD ensembles, which are excited by picosecond pulses (at 13.2~ns pulse separation), the average exciton population per dot is smaller than 0.15, when an average excitation density of 8~W/cm$^2$ (energy density 100~nJ/cm$^2$ per pulse) is used.~\cite{Berstermann}

\begin{figure}[t]
\centering
\includegraphics[width=6cm]{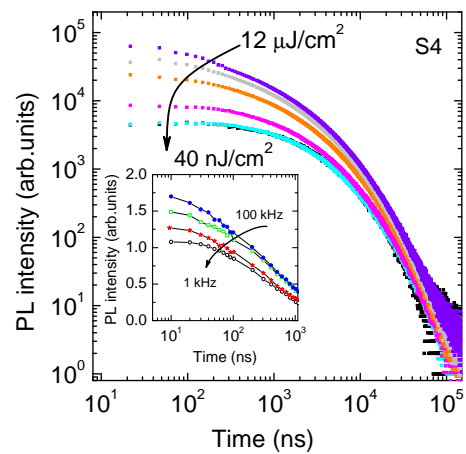}
\caption{\label{fig3} (Color online) Low temperature ($T=5$~K) PL kinetics obtained at the maximum of the IQD band (1.856 eV) for the S4 structure using different energy densities of excitation pulse (note the double-logarithmic scale). The excitation pulse ends at $t=10$~ns. The pulse repetition frequency is 1.5~kHz, which is sufficiently low to monitor a PL intensity decrease by five orders of magnitude between sequent laser pulses. From top to bottom the excitation pulse densities $P$ given in nJ/cm$^2$ are: $1.2 \times 10^4$, $4 \times 10^3$, $1.2\times 10^3$, $400$, $120$, and $40$. The inset demonstrates the PL kinetics measured at $P = 900$~nJ/cm$^2$ for different pulse repetition frequencies, from top to bottom in kHz: 100, 30, 10, and 1. While for the power density a logarithmic scale is again chosen, the intensity axis is scaled linearly, in order to underline the changes in the PL decay.}
\end{figure}

However, in indirect band gap QDs with long exciton lifetimes the optical excitation has to be carefully chosen in order to avoid accumulation of electron-hole pairs and formation of multiexciton complexes in the QDs. For that purpose, specific experimental conditions need to be established. The number of excitons, which are photogenerated in the QD surrounding matrix and captured in the QDs per laser pulse, is mainly determined by the pulse energy density, but is independent of the QD band gap structure because the relaxation from the excited $\Gamma$ to the X ground state is very fast. The repetition frequency of the excitation pulses should be reduced to a level that there is sufficient time for the excitons to recombine between subsequent pulses, so that multi-exciton complexes are not created. Since the lifetime of the indirect excitons in the (In,Al)As/AlAs QDs exceeds the one of direct excitons in (In,Ga)As/GaAs QDs by up to five orders of magnitude~\cite{Shamirzaevb}, the pulse repetition frequency as well as the average excitation density should be decreased correspondingly. To ensure that we actually study the recombination dynamics of single excitons in the (In,Al)As/AlAs QDs, the PL kinetics was measured at different excitation densities and repetition frequencies.

\begin{figure}[t]
\centering
\includegraphics[width=8cm]{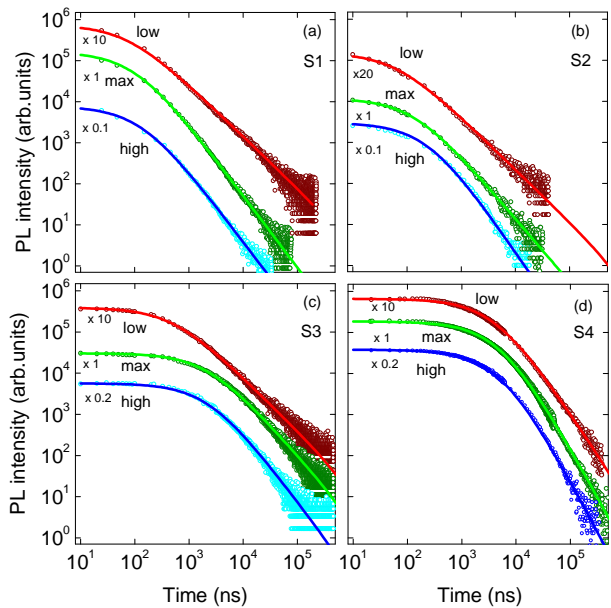}
\caption{\label{fig4} (Color online) Low temperature ($T=5$~K) PL kinetics measured for the structures S1-S4 at the intensity maximum ('max') and at half of this maximum on either the high energy ('high') or the low energy ('low') side of the IQD emission band. The excitation density amounts to $P= 40$~nJ/cm$^2$ and the pulse repetition frequency is 1.5~kHz. The excitation pulse ends at 10~ns. The scaling factors of the PL intensities are introduced for better visualization. The thick solid lines show the modeled results with the distribution of $G(\tau)$ described by Eq.~(3) with the parameters presented in Table~\ref{table}. For the S2 structure curve 'low' corresponds to the exciton recombination in QDs with typical diameter $D^{*}_{\text{L}}$.}
\end{figure}

Figure ~\ref{fig3} shows low temperature PL kinetics measured at the IQD band maximum of the S4 structure for different excitation pulse energy densities. The kinetics of the other structures is similar to the presented one. The transient PL data are plotted on a double logarithmic scale, which is convenient to illustrate the nonexponential character of the decay over a wide range of times and PL intensities. The recombination kinetics demonstrates two distinctive stages: (i) a relatively flat PL decay immediately after the excitation pulse up to approximately 1~$\mu$s and, subsequently, (ii) a reduction in the PL intensity which can be described by a power-law function $I(t) \sim (1/t)^{\alpha}$ as shown in our previous studies.~\cite{Shamirzaeva} One can see that in the case of the stage (i), a high power excitation results in a fast decay of the exciton PL. It can be assigned to the recombination of multi-exciton complexes.~\cite{Zwiller} By decreasing the power down to $P = 120$~nJ/cm$^2$ the decay decelerates, below this power the intensity does not temporally change thus indicating a saturation level. It is induced by the recombination of single excitons in the QDs. Taking into account the absorption of the laser light in the AlAs matrix~\cite{Monemar, Casey} together with the QD density, the average number of excitons captured in a QD per pulse is estimated to about 0.3 for $P = 120$~nJ/cm$^2$. An increase in the repetition rate of the excitation pulses at fixed pulse power also results in an acceleration of the initial kinetics stage, as shown in the inset of Fig.~\ref{fig3}. As the sample has to be excited by each pulse when it has reached its equilibrium state, for the further studies presented in this paper we select $P = 40$~nJ/cm$^2$. It corresponds to an average QD exciton-population of 0.1 per pulse at a repetition frequency of 1.5~kHz which equals to a time interval of 670~$\mu$s between the sequent pulses.

Fig.~\ref{fig4} shows PL decays for the structures S1-S4 which have been measured at different energies. The selection of different detection energies provides information on the exciton recombination in QDs with different characteristic sizes in the ensemble. Hereby, the energy of the intensity maximum of the IQD band (the curves labelled 'max' corresponding to recombination in QDs with the diameter $D_{\text{AV}}$) and of the half maximum (curves 'high' and 'low' related to recombination in QDs with diameters $D_{\text{S}}$ and $D_{\text{L}}$, respectively). The following features of the PL decay can be extracted: (i) for each structure the exponent ${\alpha}$ of the power-law decay is determined by fitting of the second stage of the decay curves with the form $I(t) \sim (1/t)^{\alpha}$, as listed in Table~\ref{table}. The exponent $\alpha$ increases monotonically across the IQD band from the low to the high energy side. (ii) The second decay stage starts hundred nanoseconds after the end of the excitation pulse in the unannealed structures S1 and S2, while for the annealed structures S3 and S4 it begins several microseconds after the pulse. The PL decay obviously depends on the sample characteristics and as it will be shown in the following, it is affected by both the typical QD size in the ensemble and the interface sharpness.

To obtain a quantitative description of the effect of QD size and interface sharpness on the exciton lifetime in the (In,Al)As/AlAs QD ensembles, we need to construct the distribution function  $G(\tau)$, which controls the observed PL kinetics.

\subsection{Exciton-lifetime distribution in (In,Al)As/AlAs QD ensembles}
Nonexponential decays of the exciton PL intensity $I(t)$ are frequently described by stretched-exponentials of the form $I(t) \propto \exp[-(t/\tau)^{\beta}]$, including a constant lifetime $\tau$ and a dispersion factor $\beta$.~\cite{Phillips, Guillois} The stretch parameter $0<\beta\leq 1$ qualitatively expresses the underlying distribution function $G(\tau)$: a broad distribution results in $\beta \ll 1$, while for a narrow one $\beta$ is about $1$. However, the evaluation of the lifetime distribution on the basis of the stretched-exponential model is mathematically complicated and feasible only for specific $\beta$ values (see Ref.~\onlinecite{Driel} and references therein). Alternatively, the distribution $G(\tau)$ can be determined using the following equation:
\begin{equation} \label{eq1}
I(t) = \int_0^{\infty} G(\tau) \exp \left( -\frac{t}{\tau} \right ) dt .
\end{equation}
Here, $G(\tau)$ is established via either the numerical solution of the integral equation (\ref{eq1}),~\cite{Siemiarczuk,Kapitonov,Delerue} or an assumed analytical expression of $G(\tau)$ with a set of fitting parameters. Van Driel et al.~\cite{Driel} assumed recently that the most successful distribution function to model $G(\tau)$ in a QD ensemble among different analytical expressions like normal, Lorentzian, etc distributions is a log-normal function given by:
\begin{equation} \label{eq2}
G(\tau) = \frac{A}{\tau^2} \exp \left( -\frac{\ln(\tau_0/ \tau)}{w} \right )^2 ,
\end{equation}
with the constant $A$ and the maximum $\tau_0$ of the exciton lifetime distribution. The dimensionless parameter $w$ describes the distribution width $\Delta_{1/\tau}$ of the inverse recombination times at the $1/e$ level: $ \Delta_{1/\tau} = \frac{2}{\tau_0} \sinh(w)$. This distribution was successfully used to describe the nonexponential decay of the exciton PL intensity over 2-3 orders of magnitude for different QD systems with a continuous distribution of direct exciton lifetimes. Among them are ensembles of CdSe/ZnSe colloidal QDs~\cite{Nikolaev} and dye molecules embedded in a photonic crystal.~\cite{Vallee}

\begin{figure}[t]
\centering
\includegraphics[width=6 cm]{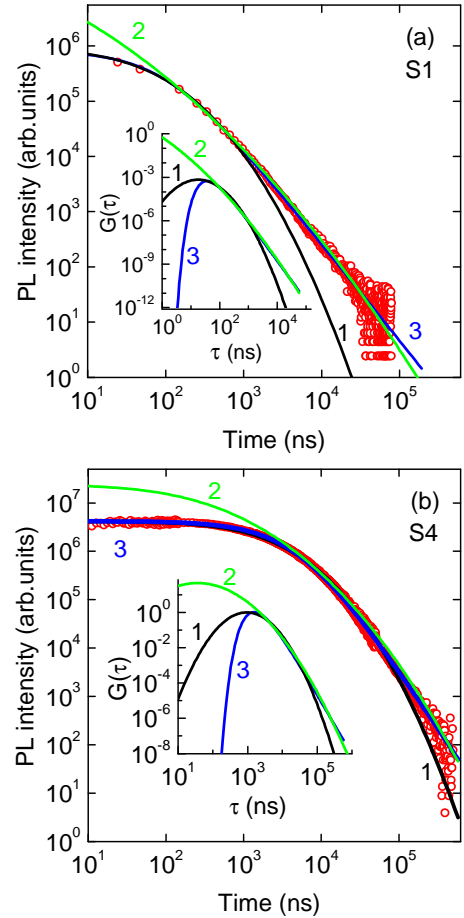}
\caption{\label{fig5} (Color online) PL decay of the samples S1 and S4 of the (In,Al)As/AlAs QDs measured at the IQD band maximum for $T = 5$~K, the experimental data are shown by circles. The excitation density is $P = 40$~nJ/cm$^2$, and the pulse repetition frequency is 1.5~kHz. The end of the excitation pulse corresponds to the time 10~ns. The modeling of the PL decays by different distribution functions $G(\tau)$ is shown by the lines. The corresponding distribution functions are given in the insets. (a) Structure S1: curves '1' and '2' are based on model calculations using the log-normal distribution $G(\tau)$ of Eq.~(\ref{eq2}) with parameter sets ($\tau_{0}=0.2$~$\mu$s, $w=1.55$) and ($\tau_{0}=0.05$~$\mu$s, $w=3.50$), respectively. Curve '3' is for $G(\tau)$ of Eq.~(\ref{eq3}) with the parameters $\tau_{0}=0.1$~$\mu$s, and $\gamma=2.75$. (b) Structure S4: curves '1' and '2' are also modeled by $G(\tau)$ of Eq.~(\ref{eq2}) with the parameter sets ($\tau_{0}=5.8$~$\mu$s, $w=1.35$) and ($\tau_{0}=3.0$~$\mu$s, $w=2.10$), respectively. Curve '3' is given by $G(\tau)$ with $\tau_{0}=5.2$~$\mu$s, and $\gamma=3.40$.}
\end{figure}

Since in (In,Al)As/AlAs QDs the excitons recombine via radiative recombination only,~\cite{note2} the nonexponential decay is the result of the dispersion of the excitonic radiative times in the ensemble. In order to determine $G(\tau)$ we follow the approach of Ref.~\onlinecite{Driel} and fit the PL kinetics with Eq.~(\ref{eq1}) using the log-normal distribution of Eq.~(\ref{eq2}). Unfortunately, the log-normal distribution does not allow us to describe the PL decay satisfactorily over the whole dynamical range. One can see in Fig.~\ref{fig5} for the structures S1 and S4 that for the decay curves, measured at maximum intensity of the IQD band (corresponding to recombination in QDs with diameter $D_{\text{AV}}$), this distribution allows to fit either the initial stage (curves '1') or the long-time stage (curves '2') of the decay, using different sets of parameters which are given in the figure caption.

In order to describe the exciton PL decay in our structures over the whole dynamical range of five orders of magnitude we propose a non-symmetric phenomenological distribution $G(\tau)$, which is suitable for fitting power-law decays $I(t) \sim (1/t)^{\alpha}$:
\begin{equation} \label{eq3}
G(\tau) = \frac{C}{\tau^{\gamma}} \textnormal{exp}\left[-\frac{\tau_{0}}{\tau}\right] .
\end{equation}
Here $C$ is a constant, and $\tau_0$ characterizes the maximum of the distribution of exciton lifetimes. The parameter $\gamma$ in the above Eq.~(\ref{eq3}) is defined as $\alpha+1$. By use of a double-logarithmic scale the power-law decay $(1/t)^{\alpha}$ represents a line with slope ${\alpha}$. Note, in knowledge of the exponent $\alpha$ of the experimental decay curve, only a single free parameter $\tau_0$ is required to describe the PL decay by Eq.~(\ref{eq3}).

We use Eq.~(\ref{eq3}) to fit the decay curves in Fig.~\ref{fig5}. The experimentally determined ${\alpha}$ values (see Table~\ref{table}) yield the values of $\gamma$ = 2.75 and 3.40 for the structures S1 and S4, respectively. One can see excellent agreement between the experimental data and the calculations for the whole PL decay comprising five orders of magnitude from the low nanosecond to the high microsecond region, shown by the curves '3' in Figs.~\ref{fig5}(a) and ~\ref{fig5}(b). The fits were obtained for the $G(\tau)$ of Eq.~(\ref{eq3}), depicted in the insets of Fig.~\ref{fig5} by the curves labeled '3'. The used fit parameters are $\tau_0 = 0.1$~$\mu$s and $\tau_0 = 5.2$~$\mu$s for the structures S1 and S4.

\begin{figure}[b]
\centering
\includegraphics[width=7cm]{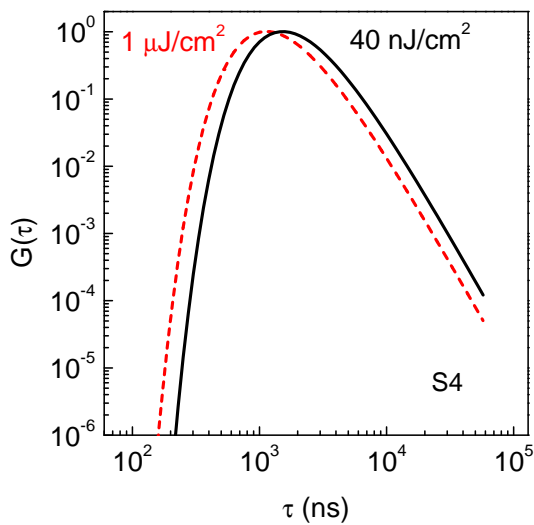}
\caption{\label{fig6} (Color online) Normalized distribution functions $G(\tau)$ corresponding to QD ensembles with the same average diameter $D_{\text{AV}} = 19.8$~nm (structure S4). The two dependences are obtained by fitting the recombination kinetics measured for different excitation energy densities: 40 nJ/cm$^2$ (black solid line) and 1~$\mu$J/cm$^2$ (red dashed line).}
\end{figure}

Let us discuss now the parameter $\gamma$ in Eq.~(\ref{eq3}). The comparison of the distribution functions Eqs.~(\ref{eq2}) and (\ref{eq3}), as shown by the curves '1' and '3' in the insets to Fig.~\ref{fig5}, clearly demonstrates that the strong difference in $G(\tau)$ for lifetimes smaller than $\tau_0$ has very little influence on the initial stage of the PL decay ($ \tau_0 \geq t$). Thus, the decay curves are mainly contributed by recombination of excitons with lifetimes exceeding $\tau_0$. Therefore, the value of the parameter $\gamma$, which is the exponent of the long-lifetime tail of $G(\tau)$, can be used as qualitative measure for the effective width of the distribution $G(\tau)$. According to Eq.~(\ref{eq3}), an increase in $\gamma$ reduces the dispersion of the $\tau$ values which contribute to the long-time tail of the kinetics curve.

We would also like to demonstrate that the filling of the QDs with multi-exciton complexes at high excitation densities distorts $G(\tau)$. The Fig.~\ref{fig6} illustrates the $G(\tau)$ distributions obtained via fitting of the recombination kinetics for the exciton population of less than one (40~nJ/cm$^2$), and in case of multi-exciton occupation (1~$\mu$J/cm$^2$). One can clearly see in Fig.~\ref{fig6} that multi-exciton QD occupation shifts the distribution maximum to shorter times. Thus, in order to reveal the $G(\tau)$, being intrinsic for the ensemble of indirect band gap QDs, the average exciton population in the dots should be less than unity.

\subsection{Effect of interface sharpness on exciton lifetime}
Using Eqs.~(\ref{eq1}) and (\ref{eq3}) we determine the $G(\tau)$ distributions for the four studied structures by fitting the decay curves presented in Fig.~\ref{fig4}. The values of the parameters $\gamma = \alpha+1$, and $\tau_0$ resulting from the fitting are collected in Table~\ref{table}. The results of the fitting are shown in Fig.~\ref{fig4} by solid lines.

\begin{figure}[b]
\centering
\includegraphics[width=7cm]{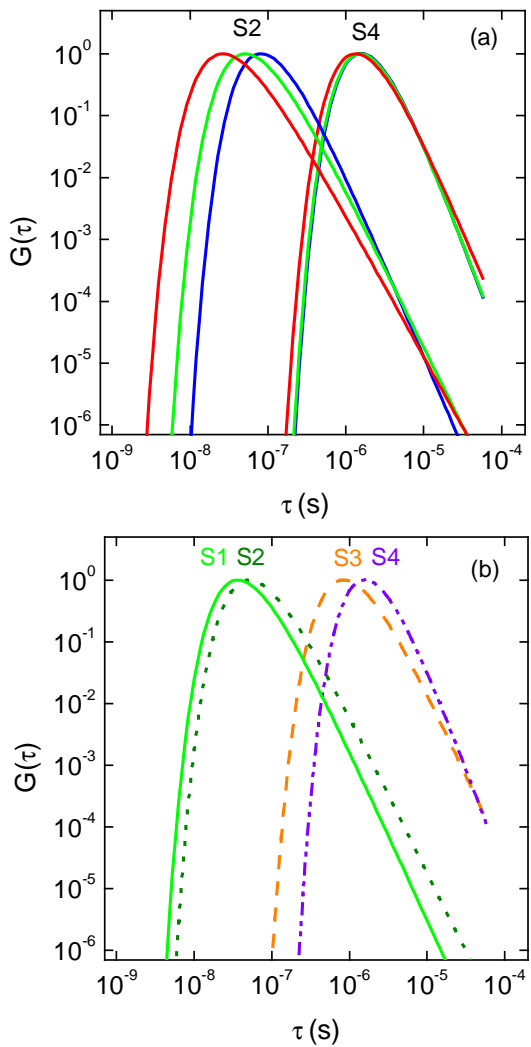}
\caption{\label{fig7} (Color online) Distributions $G(\tau)$ corresponding to (a) the QDs with diameters $D_{\text{L}} = 17$~nm, $D_{\text{AV}} = 13.8$~nm, and $D_{\text{S}} = 9$~nm for structure S2 (from left to right); and the QDs with typical diameters $D_{\text{L}} = 28$~nm, $D_{\text{AV}} = 19.8$~nm, and $D_{\text{S}}= 12$~nm for structure S4; (b) the QDs with $D_{\text{AV}}$ diameter $5.5$~nm in structure S1, $13.8$~nm in structure S2, $18.3$~nm in structure S3, and $19.6$~nm in structure S4.}
\end{figure}

The following conclusions on the distribution properties can be drawn from the data in Table~\ref{table}:
\begin{itemize}
\item A monotonic decrease of $\tau_0$ and $\gamma$ with increasing QD diameter (from $D_{\text{S}}$ to $D_{\text{L}}$) is a common feature for each structure.
\item The relative change of $\tau_0$ and $\gamma$ with changing QD diameter from $D_{\text{S}}$ to $D_{\text{L}}$ is larger in the as-grown structures S1 and S2 than in the annealed structures S3 and S4. This is also evidenced by Fig.~\ref{fig7}~(a) for the structures S2 (with smaller size dispersion $S_D = 60{\%} $) and S4 (with larger size dispersion $S_D = 75{\%} $).
\item Despite of the larger QD diameters in the annealed structures S3 and S4 than in the as-grown structures S1 and S2, $\tau_0$ is much smaller in S1 and S2 than in S3 and S4. Also, it is independent of the QD diameter, as demonstrated in Fig.~\ref{fig7}~(b) for the QDs with their average diameters $D_{\text{AV}}$ in structure S1 to S4.
\end{itemize}

These features give us the possibility to distinguish between the effect of QD size and interface sharpness on the exciton lifetime. One can see in Fig.~\ref{fig7}~(b) and in Table~\ref{table} that, despite of the large difference in QD diameter ($D_{\text{AV}} =5.5$ and 13.8~nm for S1 and S2), these structures have similar distributions of exciton lifetime. On the other hand, the comparison of $G(\tau)$ in QD ensembles with similar diameters ($D_{\text{L}}$= 17~nm, $D_{\text{AV}}$ = 18.3~nm, and $D_{\text{AV}}$ = 19.6~nm for the structures S2, S3, and S4, respectively) highlights a strong increase of the exciton lifetime with decreasing interface sharpness by about two orders of magnitude. Thus, the exciton lifetime in indirect band gap (In,Al)As/AlAs QDs is mainly determined by the interface sharpness, while its dependence on the QD size is much weaker. Nevertheless, the monotonic decrease of $\tau_0$ with increasing QD diameter from $D_{\text{S}}$ to $D_{\text{L}}$ for each of the studied structures indicates that effect of QDs size on recombination time is also important.

Our theoretical estimations are in good accordance with the experimental data. As demonstrated in the Appendix, the lifetime of an indirect in momentum space exciton can be described by the form $\tau \propto \exp(d/a+d/L)$ due to momentum scattering at the interface. Here, $a$ is the lattice constant, $L$ is the QD height, and $d$ is the thickness of the diffused (In,Al)As layer at the QD/matrix interface.

Note, the ratio $d/a \geq 1$ increases with the increase of the thickness $d$ of the diffused layer. Therefore, the exciton lifetime is mainly determined by $d/a$. It is reasonable to assume that $d/a$ changes weakly with the QD size for a particular structure type. Hence, the dependence of the exciton lifetime for such a structure type on the QD size is specified by the second term $d/L < 1$. As a result, the exciton lifetime decreases in agreement with the experimental data, for example for the S2 structure in Fig.~\ref{fig7}~(a).

Decrease of exciton lifetime with increase in $d/a$ value is restricted by 
a rate of phonon emission. When the rate of non-phonon recombination 
becomes smaller than that of the phonon emission the exciton recombination 
is mainly determined by the phonon emission.  Actually, we demonstrate in Ref.~\onlinecite{Shamirzaeve} that 
very high annealing temperature (950$^\circ$C), which leads to very smooth 
QD/matrix interface, results in appearance of phonon replicas in the PL 
spectrum of the annealed structures with In(Al)As/AlAs QDs.

Let us now consider qualitatively the effect of interface sharpness on the effective width of the $G(\tau)$ distribution. The exciton lifetime is proportional to $\exp(d/a+d/L)$. Therefore, the width of the lifetime distribution should be determined by variations of the argument $d/a+d/L$ value for different QDs. The dispersion of a varying quantity is inversely proportional to the square root of its mean value.~\cite{Landau} Thus, the width of $G(\tau)$ should decrease with increasing value of $d/a+d/L$. The experimental data tend to confirm this expectation. Indeed, the effective width of the $G(\tau)$ distribution (which is inversely proportional to $\gamma$) for QDs with similar sizes decreases with increasing thickness of the diffused layer. While an increase of the ratio $d/L$ with increasing QD size for a fixed ratio $d/a$ results in a reduction of $d/a+d/L$, and, thus, in an enhancement of the $G(\tau)$ width (decrease of the value $\gamma$ is shown in Table~\ref{table}).

\section{Conclusion}
The dynamics of the exciton radiative recombination in (In,Al)As/AlAs QD ensembles with a type-I band alignment, but an indirect band gap has been investigated. Due to the different dot sizes and/or QD/matrix interface sharpness the exciton recombination dynamics shows a nonexponential decay behavior that can be described by a power-law function as a result of the superposition of multiple monoexponential PL decays with different lifetimes. The lifetime of these excitons is mostly determined by the thickness of the diffused (In,Al)As layer at the QD/matrix interface, while its dependence on the QD size is weaker. We have proposed a phenomenological equation for the distribution $G(\tau)$ of the radiative exciton recombination times in such QD ensembles, which can describe the power-law PL decay over five orders of magnitude very well, using one fitting parameter only.

\section*{Acknowledgments}
We greatly acknowledge B.~Brinkmann for the help with experiment. This work was supported by the Deutsche Forschungsgemeinschaft and the Russian Foundation of Basic Research (joint grant no. 436 RUS 113/958/0-1 and RFBR grant no. 10-02-00240), by NATO CLG (grant no. 983878). Research stays of T.S.S. in Dortmund were financed by the DFG grants YA 65/14-1 and YA 65/19-1.

\section*{Appendix}
The exciton wave function can be written as
\begin{equation}\label{A1}
\phi({\bf r_{e}},{\bf r_{h}}) = f_{e}({\bf r_{e}})f_{h}({\bf r_{h}}) J({\bf r_{e}-r_{h}}),
\end{equation}
where {$\bf r_{e}$} and {$\bf r_{h}$} are the coordinates of electron and hole, $f_{e}$ and $f_{h}$ are their wave functions in absence of the electron-hole Coulomb interaction, while $J({\bf r_{e}-r_{h}})$ takes into account this interaction. The exciton recombination rate is proportional to $J^2(0) | \langle f_e \nabla f_h \rangle |^2$. If we decompose the wave functions into products of the Bloch waves times envelope wave functions $F_e({\bf r})$ and $F_h({\bf r})$ and assume that $\nabla$ acts only on the Bloch amplitudes, then we find that the exciton recombination rate is proportional to the square of module of the convolution of the envelopes:
\begin{equation}\label{A2}
\Gamma = \int F_{e}({\bf r})F_{h}^{*}({\bf r})d^{3}r.
\end{equation}
To estimate this integral we assume the following form for the envelopes in the vicinity of the interface between QD and matrix ($z=0$):
\begin{eqnarray}\label{A3}
F_{e}(z) &=& \exp (iqz) \sum_{k}A(k) \exp (ikz),\\
F_{h}(z) &=& \sum_{p}B(p) \exp (ipz),
\end{eqnarray}
where $q=\pi / a$, and $A, B$ are coefficients, which can be obtained from the boundary conditions. They are constants in the infinite crystal, but depend on the electron and hole momenta ${\bf k}$ and ${\bf p}$ in the quantum dots. 
Summing on these values spreads over $k$, $p \sim 1/L$ where $L \gg a$ is the dot size (for our lens-shaped QDs $L$ is the height of QD). Thus, we can assume $k$, $p \ll q$ and the value of the exciton recombination rate at the QD/matrix interface is determined by the integral:

\begin{equation}\label{A4}
\Gamma_{z} = \int \exp [i(p-k+q)z] dz,
\end{equation}
which is zero far from the interface, because of the oscillating factor $\exp (iqz)$. We can estimate $\Gamma_z$ at a sharp interface as
\begin{equation}\label{A5}
\Gamma_{z} = \int \exp [i(p-k+q)z] dz \sim \frac{1}{iq} \sim
\frac{ia}{\pi}.
\end{equation}
Integration over the dot interface leads to estimation of $\Gamma_z$ as the ratio of the number of atoms located at the interface to the total number of atoms in the quantum dot. Estimation of $\Gamma_z$ at a diffused interface can be done assuming that $p$ and $k$ vary smoothly with position inside the interface layer: $p(z)=\sqrt{2m_h(E-U(z))}$ and $k(z)=\sqrt{2m_e(E-U(z))}$, where $U(z)$ is the potential profile of a diffused interface. Then
\begin{equation}\label{A6}
\Gamma_{z} = \int^{+\infty}_{-\infty} \exp (i(p(z)-k(z)+q)z) dz .
\end{equation}
To estimate $ \Gamma_{z}$ we consider the integrand in Eq.~(\ref{A6}) as a function of the complex variable $z$.  We can displace the integration contour from the real axis $z$ into the upper half-plane, up to the nearest singularity $z_p$ of the potential $U(z)$. This value is about $id$, where $d$ is the characteristic thickness of the diffused interface. For
\begin{equation}\label{A7}
U(z) = \frac{U_0}{1+\exp (-z/d)}
\end{equation}
the value of $z_p$ is given by $i \pi d$; the actual value of $z_p$ depends on the model for the interface. Therefore, the integral~(\ref{A6}) contains the exponential factor $\exp(iq z_p) \sim \exp(qd)$. Evaluation of the integral similar as done for Eq.~(\ref{A6}) in Ref.~\onlinecite{Baskin} results in
\begin{equation}\label{A8}
\Gamma_{z} \sim \frac{\exp (-qd)}{q} \sim \frac{a}{\pi}\exp \left(
-\frac{d}{a} \right ).
\end{equation}
Note that estimations (\ref{A4})--(\ref{A6}) suppose the large size of the QD. Decrease of the QD size, e.g. its height, leads to increase of the electron and decrease of the hole energies due to the size quantization as well as non-zero value of the electron and hole momenta (which is about $\pi\hbar/L$).
To take this fact into account, we have to add $\pi/L$ to the exponent Eq.~(\ref{A6}), i.e. replace $q$ with $q+\pi/L$ in Eq.~(\ref{A6}) and  Eq.~(\ref{A8}). Then Eq.~(\ref{A8}) accepts the form
\begin{equation}\label{A8a}
    \Gamma_z\sim\frac{a}{\pi}\exp{\left[ -\frac{d}{a}\left(1+\frac{a}{L}\right) \right ]}.
\end{equation}
Taken into account that exciton recombination rate is inversely proportional to exciton lifetime our estimation demonstrates that $\tau \sim \exp{\left[ \frac{d}{a}+\frac{d}{L}\right ]}$  
i.e. increase in thickness of diffused layer at the QD/matrix interface results in exponential increase of exciton lifetime.


\begin{thebibliography}{}
\bibitem{note}
The recombination energy of an exciton in a QD is determined by the QD size, while photons at a certain spectral energy are contributed by QDs with similar sizes. Thus, the PL band maximum corresponds to recombination in QDs with the size given by the average QD diameter. In Ref.~\onlinecite{Shamirzaevd} we calculated confined energy levels of QDs with different diameters as function of the QD composition. The comparison of the measured energy of the PL maximum with the calculated optical transition energies in QDs with known average sizes allows us to determine the average composition of these QDs. The InAs fraction in the (In,Al)As QDs as function of the substrate temperature and the growth-interruption time is given in Ref.~\onlinecite{Shamirzaevd}.
\bibitem{note2} The long exciton lifetimes in the (In,Al)As/AlAs QDs evidence that carriers captured in QDs recombine through photon emission only. Actually, we have demonstrated in Ref.~\onlinecite{Shamirzaevh} that even a small fraction of QDs containing nonradiative centers of 5$\%$ would only decrease the PL decay time down to about 1~{$\mu$}s due to the long-distant transfer of the exciton energy to the nonradiative centers.
\end{thebibliography}
\end{document}